\documentclass[letterpaper, aps, prd, twocolumn, superscriptaddress, showpacs, nofootinbib]{revtex4}
\pdfoutput=1

\usepackage{graphicx}
\usepackage{bm}
\usepackage{dcolumn}
\usepackage{color}
\usepackage[sort&compress]{natbib}
\usepackage[latin9]{inputenc}
\usepackage{url}
\usepackage{subfigure}
\usepackage{float}
\usepackage{longtable}
\usepackage{amssymb}
\usepackage{amsmath}
\usepackage{amsfonts}
\usepackage{array}
\usepackage{graphicx}
\usepackage{multirow}
\usepackage{xcolor}
\usepackage{ulem}
\newcommand\MyBox[2]{
  \fbox{\lower0.75cm
    \vbox to 1.2cm{\vfil
      \hbox to 1.2cm{\hfil\parbox{1.0cm}{#1\\#2}\hfil}
      \vfil}%
  }%
}

\DeclareMathOperator*{\argmin}{arg\,min}

\begin{document}

\title{Determining the core-collapse supernova explosion mechanism with current and future gravitational-wave observatories}

\newcommand*{\swin}{Centre for Astrophysics and Supercomputing, Swinburne University of Technology, Hawthorn, VIC 3122, Australia.}
\affiliation{\swin}

\newcommand*{\ozgrav}{ARC Centre of Excellence for Gravitational Wave Discovery (OzGrav), Melbourne, Australia.}
\affiliation{\ozgrav}

\newcommand*{\roma}{Universita di Roma Tor Vergata, I-00133 Roma, Italy.}
\affiliation{\roma}

\newcommand*{\valencia}{Departamento de Astronom\'{\i}a y Astrof\'{\i}sica, Universitat de Val\`encia, Dr.~Moliner 50,  46100 Burjassot (Val\`encia), Spain.}
\affiliation{\valencia}

\newcommand*{\monash}{School of Physics and Astronomy, Monash University, Clayton, VIC 3800 Australia.}
\affiliation{\monash}

\newcommand*{\ego}{European Gravitational Observatory (EGO), I-56021 Cascina, Pisa, Italy.}
\affiliation{\ego}

\newcommand*{\sns}{Scuola Normale Superiore, Piazza dei Cavalieri 7, I-56126 Pisa, Italy.}
\affiliation{\sns}

\newcommand*{\OAUV}{Observatori Astron\`omic, Universitat de Val\`encia, Catedr\'atico Jos\'e Beltr\'an 2, 46980 Paterna (Val\`encia), Spain}
\affiliation{\OAUV}

\author{Jade Powell}  \affiliation{\swin} \affiliation{\ozgrav}
\author{Alberto Iess} \affiliation{\roma}	
\author{Miquel Llorens-Monteagudo} \affiliation{\valencia}
\author{Martin Obergaulinger} \affiliation{\valencia}
\author{Bernhard M\"uller} \affiliation{\monash}
\author{Alejandro Torres-Forn\'e} \affiliation{\valencia}
\author{Elena Cuoco} \affiliation{\ego} \affiliation{\sns}
\author{Jos\'e A.~Font} \affiliation{\valencia} \affiliation{\OAUV}

\date{\today}

\begin{abstract} 

Gravitational waves are emitted from deep within a core-collapse supernova, which may enable us to determine the mechanism of the explosion from a gravitational-wave detection. Previous studies suggested that it is possible to determine if the explosion mechanism is neutrino-driven or magneto-rotationally powered from the gravitational-wave signal. However, long duration magneto-rotational waveforms, that cover the full explosion phase, were not available during the time of previous studies, and explosions were just assumed to be magneto-rotationally driven if the model was rapidly rotating. Therefore, we perform an updated study using new 3D long-duration magneto-rotational core-collapse supernova waveforms that cover the full explosion phase, injected into noise for the Advanced LIGO, Einstein Telescope and NEMO gravitational-wave detectors. We also include a category for failed explosions in our signal classification results. We then determine the explosion mechanism of the signals using three different methods: Bayesian model selection, dictionary learning, and convolutional neural networks. The three different methods are able to distinguish between neutrino-driven explosions and magneto-rotational explosions, even if the neutrino-driven explosion model is rapidly rotating. However they can only distinguish between the non-exploding and neutrino-driven explosions for signals with a high signal to noise ratio.

\end{abstract}

\maketitle
\section{Introduction}
\label{sec:intro}

In the last few years, current ground-based gravitational-wave detectors Advanced LIGO \cite{2015CQGra..32g4001L}, Advanced Virgo \cite{2015CQGra..32b4001A}, and KAGRA \cite{2020arXiv200505574A} have reached the sensitivities required for the detection of gravitational waves from merging binary black holes and neutron stars. A small population of compact binary sources have now been detected \cite{2019PhRvX...9c1040A, 2021PhRvX..11b1053A, 2021arXiv211103606T}. However, there are many other potential sources for ground-based, gravitational-wave detectors~\cite{2020LRR....23....3A}. One of the most promising is core-collapse supernovae (CCSNe) \cite{2016PhRvD..93d2002G, 2021PhRvD.104j2002S}. To date, the optically-targeted searches from the LIGO-Virgo-KAGRA Collaborations have not made any detections of gravitational waves from a CCSN \cite{2020PhRvD.101h4002A, 2023arXiv230516146S}. The sensitivity to these sources will improve in future gravitational-wave detectors, such as the proposed high frequency detector NEMO \cite{2020PASA...37...47A}, the Einstein Telescope (ET) \cite{2011CQGra..28i4013H}, and Cosmic Explorer \cite{2021arXiv210909882E}.  

Determining the astrophysical properties of CCSNe from their gravitational-wave emission is difficult, as CCSN signals have stochastic components, and generating waveforms from numerical simulations is computationally expensive. However, there are clear deterministic aspects of the signals that have enabled studies to ascertain the astrophysical parameters of the gravitational-wave source.
Previous work on the estimation of CCSN parameters has focused on the inference of properties such as the equation of state or the rotation rate \cite{2009PhRvD..80j2004R, 2014PhRvD..90d4001A, 2014PhRvD..90l4026E, 2021PhRvD.103b4025E,2023arXiv230803456P}, or the mass and radius of the proto-neutron star (PNS) \cite{2021PhRvD.103f3006B, 2022PhRvD.105f3018P, 2023PhRvD.107h3029B}. There have also been many efforts to try and determine the CCSN explosion mechanism from the gravitational-wave emission~\cite{2012PhRvD..86d4023L,2016PhRvD..94l3012P,2017PhRvD..96l3013P,2022MNRAS.512.3815S}. 

Stars with a zero age main sequence (ZAMS) mass larger than about $8\,M_{\odot}$ are expected to end their life as CCSNe. Nuclear burning in the stellar core stops when it consists of iron nuclei. The core will then collapse until it exceeds
 nuclear saturation density. The stiffening of the equation of state due to repulsive nuclear forces leads to a rebound of the core, and a shock wave is launched out from the core. The shock quickly loses energy and stalls and must be revived by depositing energy in the post-shock region to make it propagate outward dynamically and power a CCSN explosion. The method in which the shock wave gains energy to power the full explosion is known as the CCSN explosion mechanism. 
Numerical simulations suggest several possible CCSN explosion mechanisms 
\cite{2015PASA...32....9F, 2021Natur.589...29B, 2012ARNPS..62..407J}. 
 
The majority of CCSNe are thought to explode by the neutrino-driven explosion mechanism \cite{2017hsn..book.1095J}. Neutrinos carry most of the energy in a CCSN explosion. In the neutrino-driven explosion mechanism, a very small fraction of the neutrino energy is reabsorbed behind the shock to power the explosion. The gravitational-wave signals from stars that exploded by this mechanism contain most of their energy in f- or g-modes whose frequencies depend on the mass and radius of the PNS and are in general above $\sim 500$\,Hz \cite{2019MNRAS.487.1178P, 2019ApJ...876L...9R, 2018ApJ...865...81O, 2018ApJ...857...13P, 2020PhRvD.102b3027M}. They may also contain low-frequency (below $\sim 500$\,Hz) modes due to the standing accretion shock instability (SASI) \cite{2017ApJ...835..170B, 2023PhRvD.107f3014W}. The SASI typically stops after the shock is revived, and is therefore usually a more prominent feature before the shock revival time in exploding models, or for a prolonged time  
in failed explosion models. 

Rapidly-rotating CCSNe with powerful magnetic fields may explode by the magneto-rotational explosion mechanism \cite{2018PAN....81..266B, 2020MNRAS.492.4613O, 2023MNRAS.522.6070P}. This mechanism produces more powerful explosions that may be associated with hypernovae or gamma-ray bursts. The rotation produces a large broadband spike in the gravitational-wave signal amplitude at the time of the core bounce. The rotation and magnetic fields can also change the relationship between the f-/g-modes and the mass and radius of the PNS \cite{2020MNRAS.494.4665P, 2021ApJ...914...80P}. Other explosion mechanisms may exist, such as explosions powered by a hadron quark phase transition \cite{2020ApJ...894....9F, 2022PhRvD.106l3037Z, 2023arXiv230106515J}, 
however we do not consider those potential mechanisms in this work. 

In \cite{2009CQGra..26j5005H}, the authors first showed how to distinguish between two different supernova explosion models using principal component analysis (PCA). The authors in \cite{2012PhRvD..86d4023L} then extended this work to combine PCA and Bayesian model selection to distinguish between the neutrino-driven, magneto-rotational and acoustic explosion mechanisms.  
They used a single Advanced LIGO detector and waveforms from 2D CCSN simulations. This work was further extended in \cite{2016PhRvD..94l3012P} to include a network of gravitational-wave detectors to analyse the same set of waveforms. However, there can be significant differences between CCSN waveforms when computed using 2D or 3D simulations. Therefore in \cite{2017PhRvD..96l3013P}, the authors updated the study in \cite{2016PhRvD..94l3012P} to incorporate waveforms from 3D simulations that include both gravitational-wave polarisations. How well we can determine the explosion mechanism in the next generation of gravitational-wave detectors was explored in \cite{2019PhRvD..99f3018R}. 

The potential to infer the explosion mechanism has also been explored using machine learning techniques. Several studies have shown how to distinguish between different types of CCSN signals using neural networks \cite{2020PhRvD.102d3022C, Iess_2020}. Another promising technique is dictionary learning. This approach was applied in \cite{2022MNRAS.tmp..710S} to distinguish between the neutrino and magneto-rotational explosion mechanisms using 2D waveforms and a single gravitational-wave detector. 

Although inferring the CCSN explosion mechanism from a gravitational-wave signal has already been extensively studied, there is still further improvements needed to be prepared for a real CCSN gravitational-wave event. The biggest outstanding issue with previous studies is the lack of waveforms available for the magneto-rotational explosion mechanism. Those studies used waveforms that only contain the core-bounce signal and were stopped before they reached the time of the explosion. It is not possible to know that the explosion would have truly been magneto-rotationally powered if the simulation is stopped before the shock revival time. Moreover, gravitational-wave signals from stars that are rapidly rotating, but still undergo neutrino-driven explosions, are very similar to gravitational-wave signals from magneto-rotational explosions \cite{2020MNRAS.494.4665P}. Having more complete waveforms will make distinguishing the two explosion mechanisms much more difficult than in previous studies that only used very short duration rapidly-rotating waveforms to represent the magneto-rotational mechanism, and only used non-rotating waveforms for the neutrino-driven explosions. 
Previous studies also used slightly different data, making it difficult to carry out a direct comparison between the different proposed classification methods. 

Therefore, in this study, we use long-duration 3D magneto-rotational explosion mechanism waveforms that have recently become available, where we define a magneto-rotational explosion model as a model where the magnetic fields play either a supporting or leading role in the shock revival. We also include a non-exploding class, as stars that do not undergo shock revival still emit gravitational waves before black hole formation. All of our non-exploding models were non-rotating. We also include rotating models in our neutrino-driven explosion class, as they should be more difficult to distinguish from the magneto-rotational explosion waveforms. We also carry out the first direct comparison between three of the previously used methods: Bayesian model selection, convolutional neural networks, and dictionary learning. Moreover, we test our methods using three different gravitational-wave detectors, which are Advanced LIGO, Einstein Telescope and the proposed Australian high-frequency detector NEMO. 
Our results show that we can distinguish between neutrino-driven explosions and magneto-rotational explosions, even when neutrino-driven explosions occur in rotating progenitors. However, our methods struggle to distinguish between neutrino-driven explosions and non-exploding models, as the prolonged SASI component in non-exploding models is only visible at high signal to noise ratios (SNR). 
 
The paper is structured as follows: In Section \ref{sec:waveforms}, we describe all of the CCSN waveforms used in this study, and explain the differences between the various explosion mechanisms. In Section \ref{sec:codes}, we give a brief description of the Bayesian model selection, dictionary learning and convolutional neural network methods we use to determine the explosion mechanism. In Section \ref{sec:noise}, we describe how we produce the noise and signals for the different gravitational-wave detectors. In Section \ref{sec:results}, we describe the results for all three methods, and a discussion and conclusions are given in Section \ref{sec:conclusion}. 

\section{Supernova waveforms}
\label{sec:waveforms}

\begin{figure*}
\includegraphics[width=16cm]{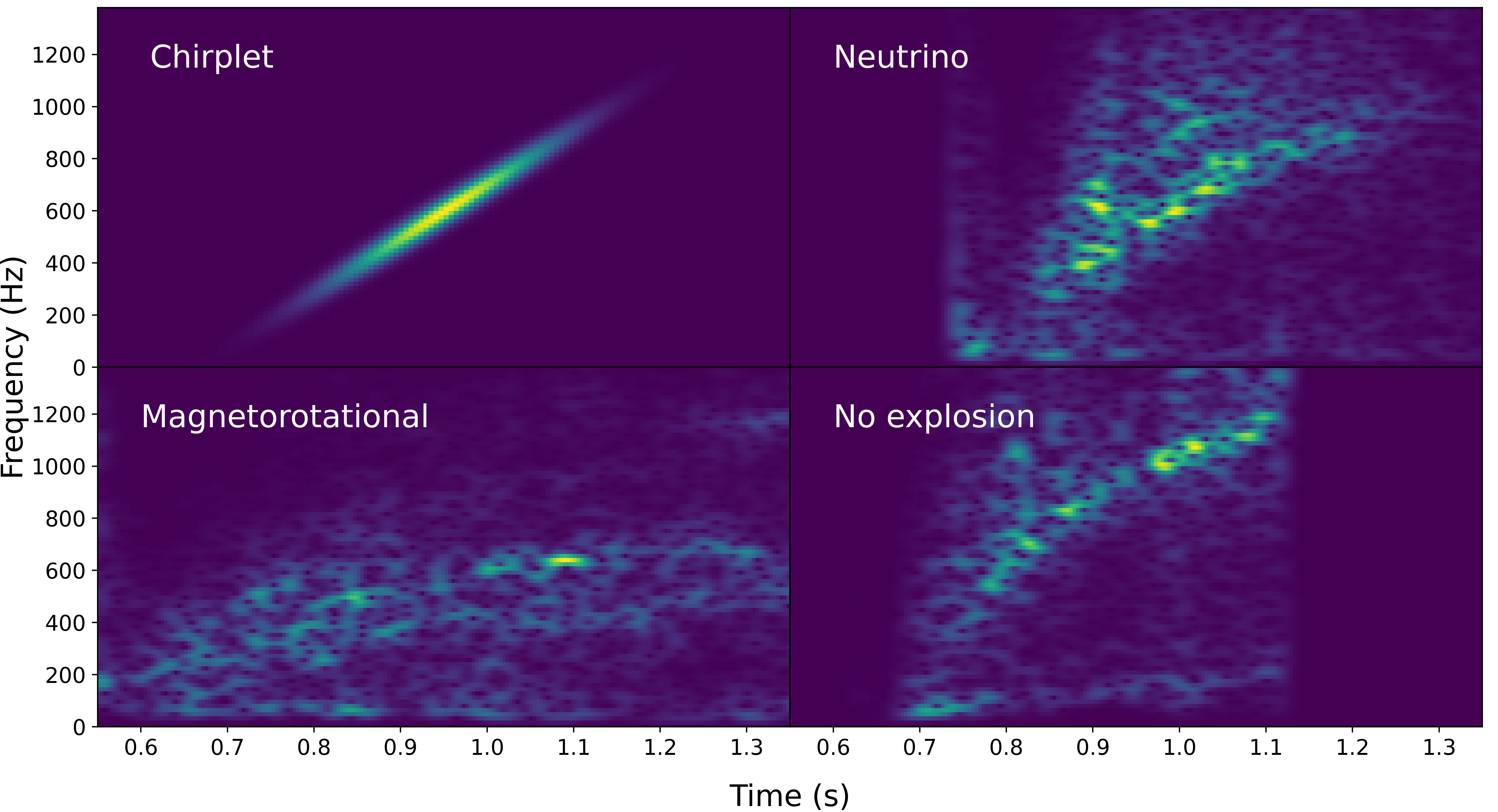}
\caption{An example of each of the four types of CCSN waveforms used in this study, shown with time-frequency plots (or spectrograms). Top left is an example chirplet from the training set with a frequency of $600$\,Hz and quality factor $Q$ 300. Top right is the neutrino driven explosion mechanism waveform model \texttt{y20} from \cite{2020MNRAS.494.4665P}.  
Bottom left is the magneto-rotational mechanism waveform model \texttt{A39} from \cite{2020MNRAS.492...58B}. Bottom right is the non-exploding waveform model mesa20\_pert from \cite{2018ApJ...865...81O}.  }
\label{fig:waveforms}
\end{figure*}

We use four different classes of CCSN waveforms for this project. They are non-exploding, neutrino-driven explosions and magneto-rotational explosions. We also include a chirplet class as an example of a generic phenomenological waveform for cases when the gravitational-wave signal does not match any of our proposed explosion mechanisms. An illustration of one waveform from each of the four classes is shown in Figure \ref{fig:waveforms}. We describe next which waveforms were used for training our algorithms and which waveforms were used to obtain the results. It is important to test our methods using waveforms that are not included in the training, as a real CCSN signal will never match exactly one of our simulated waveforms. 

\subsection{Non-exploding}

In the non-exploding case, the shock is not revived, and in some cases, the star can quickly form a black hole. Gravitational waves are emitted from shortly after core-bounce up to the time of black hole formation. Non-exploding models often have less gravitational-wave energy than exploding models. However, high mass stars that do not fully explode can also have fairly large gravitational-wave energies \cite{2021MNRAS.503.2108P}. Models that do not explode often have a 
more prolonged low frequency component in the gravitational-wave emission due to the SASI, in comparison to the neutrino-driven and magneto-rotational explosion models.  
The low-frequency SASI mode, which occurs below $\sim 500$\,Hz and rises in frequency with time, is clearly visible in the non-exploding example shown in the bottom-right panel of Figure \ref{fig:waveforms}. 
All of the non-exploding models used here are non-rotating. 

\subsubsection{Training}

For the non-exploding training data, we employ 7 different waveforms. For each waveform, we use the signal as measured at the equator and the pole. The first is model \texttt{s40 NR} from Pan et al.~\cite{2018ApJ...857...13P}. The model was simulated with the code \texttt{FLASH} and has a $40\,M_{\odot}$ progenitor star, 
and the waveform has a duration of 0.77\,s. We also use three models from O'Conner \& Couch \cite{2018ApJ...865...81O}. They are simulations of a $20\,M_{\odot}$ progenitor star with the SFHo equation of state (EoS), and we use the \texttt{mesa20\_gw}, \texttt{mesa20\_pert\_gw}, and \texttt{mesa20\_LR\_gw} models. The fifth non-exploding model we use is the $13\,M_{\odot}$ progenitor model from Radice et al.~\cite{2019ApJ...876L...9R}, which has the SFHo EoS. The last two models are the \texttt{z100} models from Powell et al.~\cite{2021MNRAS.503.2108P} simulated with the \texttt{CoCoNuT} code. They have the same $100\,M_{\odot}$ progenitor, but two different EoS, SFHo and SFHx.

\subsubsection{Injections}

We add two different non-exploding models into the detector noise to test our methods. The first one is the \texttt{s18np} model from Powell \& M\"uller \cite{2020MNRAS.494.4665P}, simulated with the \texttt{CoCoNuT} code. It is a non-rotating $18\,M_{\odot}$ progenitor model simulated with the LS220 EoS. Perturbations were excluded from the model to make it more difficult for the shock to revive. As a result, the model has a strong low-frequency SASI component. The waveform has a  duration of 0.56\,s. The second waveform is the \texttt{C15} model from Mezzacappa et al.~\cite{2020PhRvD.102b3027M}. It corresponds to a $20\,M_{\odot}$ non-rotating progenitor star simulated with the \texttt{Chimera} code. The model has a strong low-frequency SASI component and the highest frequency mode reaches frequencies of above 1000\,Hz.

\subsection{Neutrino-driven explosions}

Neutrino-driven explosions are thought to occur in most regular CCSNe \cite{2017hsn..book.1095J}. The most common gravitational-wave feature from stars that explode by this mechanism is the appearance of high-frequency g/f-modes. Their spectrograms may also reveal a SASI component before the shock is revived. In recent years, a large number of 3D simulations have produced waveforms for this explosion mechanism. In the following sub-sections, we list which waveforms we use for our training sets and which ones are injected into the noise. 

\subsubsection{Training}

The first waveform in the training set is the fast rotating model \texttt{s40\_FR} from Pan et al.~\cite{2018ApJ...857...13P}. It is the same progenitor model as we used in the non-exploding training set. However, the rapid rotation added to the model results in shock revival $\sim200$\,ms after core bounce. In addition, we use several neutrino-driven explosion models from Radice et al.~\cite{2019ApJ...876L...9R} with a variety of different gravitational-wave energies. They are the $10\,M_{\odot}$, $11\,M_{\odot}$, $19\,M_{\odot}$, and $60\,M_{\odot}$ progenitor models. 

The next waveform in the training set is the \texttt{y20} model from Powell \& M\"uller \cite{2020MNRAS.494.4665P}. The progenitor is a Wolf-Rayet star with a helium core mass of $20\,M_{\odot}$. The model is non-rotating and has the LS220 EoS. The simulation is 1.2\,s long and the shock is revived about 200\,ms after core bounce. The gravitational-wave emission attains a high amplitude for only the first 0.6\,s after core bounce. We also use two neutrino-driven models from Powell et al.~\cite{2021MNRAS.503.2108P}. They are the $85\,M_{\odot}$ models with the SFHo and SFHx EoS. Both models undergo a neutrino-driven explosion before forming a black hole, which results in a sharp cut off in the gravitational-wave emission. The \texttt{z85\_SFHo} model forms a black hole 0.36\,s after core bounce, and the \texttt{z85\_SFHx} model forms a black hole at 0.59\,s after core bounce. Both \texttt{z85} models have a strong SASI component in the gravitational-wave signal before the shock revival.

\subsubsection{Injections}

For the injected waveforms, we use the non-rotating $18\,M_{\odot}$ model from Powell \& M\"uller \cite{2019MNRAS.487.1178P}. It has the LS220 EoS and a duration of 0.89\,s. We also
use the first 0.54\,s of the non-rotating $12\,M_{\odot}$ model from Radice et al.~\cite{2019ApJ...876L...9R}. 
The third injected waveform is the rapidly rotating model \texttt{m39} from Powell \& M\"uller 
\cite{2020MNRAS.494.4665P}. We choose this model because rapid-rotation results in a core-bounce signal that is similar to what is observed in rapidly-rotating magneto-rotational explosions. Therefore, we expect this model to be more difficult to classify as a neutrino-driven explosion than the non-rotating neutrino-driven explosion models. 
The progenitor star has an initial helium star mass of $39\,M_{\odot}$, and an initial surface rotational velocity of $600\,\mathrm{kms}^{-1}$, and a pre-collapse core rotation rate of $0.54\,\mathrm{rad\,s}^{-1}$. The duration of the waveform is 0.98\,s. As well as the spike at core bounce, the rotation also increases the amplitude of the high frequency g-mode.

\subsection{Magneto-rotational explosions}

In the magneto-rotational explosion mechanism, the energy to revive the shock comes from the energy produced by the rotation of the PNS, which is amplified by a strong magnetic field. 
They are thought to be less common than neutrino-driven explosions. 
Very rapid rotation and strong magnetic fields are required for an explosion to be powered by this mechanism. Rotation alone significantly increases the amplitude of gravitational-wave emission. Both rotation and magnetic fields combined increase the amplitude even further, resulting in a much larger maximum detection distance than for neutrino-driven explosions. Rapid explosions mean that these signals often do not have a strong SASI component to the gravitational-wave signal. Rotation results in a spike in the time-series of the gravitational-wave emission that occurs at the core bounce time in the plus polarisation only if the viewing angle is from the equator.

\subsubsection{Training}

The first model in the training set is model \texttt{m39\_B10} from Powell \& M\"uller \citep{2023MNRAS.522.6070P}, which is a $39\,M_\odot$ model with a pre-collapse magnetic field strength of $10^{10}$\,G, and the same rotation rate as the \texttt{m39} neutrino-driven explosion model. The gravitational-wave signal is only 0.33\,s long due to a short simulation time. The shock is revived $\sim150$\,ms after the core-bounce. The gravitational-wave frequency peaks at around 1500\,Hz.

The next five models in the training set are based on a low-metallicity star with $35 \, M_{\odot}$ which at the time of collapse is rotating rapidly. The models \texttt{O}, \texttt{P}, and \texttt{W} from \cite{2020MNRAS.492.4613O} and \texttt{l1\_90d} and \texttt{l2\_gB} from \cite{2020MNRAS.492...58B} differ in the pre-collapse magnetic fields and, as a consequence, in their evolution. They all produce explosions in which the magnetic field and rotation at least play a supporting role. The initial field of model \texttt{O} is taken from the progenitor model, has an off-centre maximum field strength of $\sim 10^{12}$\,G. The core explodes at about 0.2 s after bounce. Its gravitational-wave signal has a duration of about 0.8\,s. Model \texttt{P} differs from model \texttt{O} by a multiplication of the initial poloidal field by a factor 3, while the energetically dominant toroidal component is the same. At $\sim 0.15$\,s, an explosion starts which develops bipolar jets. The gravitational-wave signal has a duration of about 1.5\,s. Instead of the progenitor field, model \texttt{W} starts with a weak dipolar initial field normalised to $10^{10}$\,G. As a consequence, the influence of the magnetic field on the dynamics is modest. An explosion driven mostly by neutrino heating starts at a similar time as in model \texttt{O}. We use about 0.8\,s of gravitational-wave data for this model. 

Models \texttt{l1\_90d} and \texttt{l2\_gB} are initialised with a dipolar field tilted by $90^{\circ}$ w.r.t.~the rotational axis, i.e., with the magnetic poles in the equatorial plane, and with a quadrupolar field aligned with the rotational axis, respectively. In both cases, the field strength is normalised to a maximum value of $10^{12}$\,G. In both models, the explosion is driven by the magnetic fields. It sets in about 0.1\,s after bounce and gives rise to jets. About 0.65\,s and 0.88\,s of the gravitational-wave signal are available for \texttt{l1\_90d} and \texttt{l2\_gB}, respectively. As all models have the same rotational profiles, the maximum amplitudes of the gravitational-wave emission at bounce are similar. 
Later on, they develop different waveforms, though usually showing features with frequencies increasing beyond 1\,kHz.

The final two models of the training set, models \texttt{A26} and \texttt{A39}, have progenitors of initial masses of $26$ and $39\, M_{\odot}$, respectively, from model series B produced by \cite{2018ApJ...858..115A} under the assumption of chemically homogeneous evolution. Both progenitor models include profiles of the rotation and magnetic field, which we directly use as initial data for the simulations. The initial models contain alternating magnetised and unmagnetised shells. In both cases, the maximum field strength ($\approx 10^{11}$\,G) is reached in small regions off-centre. Explosions set in at about 0.6\,s and 0.3\,s after bounce, respectively. They are strongly affected by rotation and the magnetic fields. We use a span of about 1.4\,s and 0.87\,s of gravitational-wave data from \texttt{A26} and \texttt{A39}, respectively. The maximum signal amplitudes are lower than for the models based on progenitor \texttt{35OC}, while the frequency range is similar.

\subsubsection{Injections}

The first model used for the injections is \texttt{m39\_B12} from Powell \& M\"uller \citep{2023MNRAS.522.6070P}. Is it also a simulation of the same $39\,M_\odot$ progenitor model, however this one has a stronger magnetic field strength of $10^{12}$\,G, but the same rapid rotation rate. The model undergoes shock revival at $\sim150$\,ms, as in model \texttt{m39\_B10}. However the gravitational-wave signal is much longer as the simulation ended 0.68\,s after the core-bounce. The model 
reaches a maximum frequency of above 2000\,Hz. 
The second model, \texttt{A13}, belongs to the series of chemically homogeneous stars of \cite{2018ApJ...858..115A}. It has an initial mass of $13\,M_{\odot}$ and a maximum initial magnetic field strength of $\approx 10^{11}$\,G. Magnetically driven shock revival sets in at $\approx 0.3$\, s after bounce and leads to jet formation. We use a gravitational-wave signal with a 1.5\,s duration. The bounce signal is at $\approx 25$\,cm weaker than in \texttt{m39\_B12}. Later on, we find modes of up to $\sim 1500$\,Hz and strong contributions to the signal below $\approx 1000$\,Hz.

\subsection{Chirplets}

As mentioned before, we add a fourth model category to show what happens when the gravitational-wave signal does not belong to any of our three potential explosion mechanisms. We choose a chirplet model, as it is similar in a spectrogram to a real CCSN signal. The equation for the chirplet is given by,
\begin{equation}
h = h_{\rm rss}  \exp( -dt^2/\tau^2)  \cos(2 \pi f dt + \pi \dot{f} dt^2 ) 
\end{equation}
where $h_{\rm rss}$ is the root sum squared amplitude, $dt$ is each time step, $f$ is the central frequency, $\dot{f}$ is the rate of change of frequency, and $\tau$ is the duration, which is defined as $\tau = Q / 2\pi f$, where $Q$ is the quality factor.

\subsubsection{Training}

For the chirplets used for the training, we use a variety of values for both frequency and quality factor. They are given in Table \ref{tab:chirps}. The frequency and $Q$ values are chosen so that they will produce gravitational-wave signals that are similar to the time-frequency morphology and duration of the other CCSN waveforms from the other explosion mechanisms. 

\begin{center}
\begin{table}
\begin{tabular}{ |c|c|c|c| } 
 \hline
model name & f\,(Hz) & Q   & $\dot{f}$\,(Hz/s) \\ \hline
chirplet 1 & 600 & 300 & 3183 \\ 
chirplet 2 & 600 & 380 & 3183 \\ 
chirplet 3 & 650 & 300 & 3183 \\
chirplet 4 & 650 & 380 & 3183 \\
chirplet 5 & 750 & 300 & 3183 \\
chirplet 6 & 750 & 380 & 3183 \\
chirplet 7 & 800 & 300 & 3183 \\ 
chirplet 8 & 800 & 380 & 3183 \\
chirplet 9 & 850 & 300 & 3183 \\
chirplet 10 & 850 & 380 & 3183 \\
 \hline
\end{tabular}
\label{tab:chirps}
\caption{The parameter values of the chirplet signals used for training the three different classification methods.}
\end{table}
\end{center}

\subsubsection{Injections}

For the chirplets injected into the data, we use the same parameters for every injected gravitational-wave signal. The injected signals have a frequency of 700.0\,Hz, an $\dot{f}$ of 3183\,Hz/s, and a $Q = 351.53$. The amplitude parameter $h_{\rm rss}$ is scaled to give the required signal to noise ratio (SNR).

\section{Pipelines}
\label{sec:codes}

We perform our analysis using three different methods. This section gives a brief description of each of them, including any changes made since previous publications.  

\subsection{Bayesian Model Selection}

The Supernova Model Evidence Extractor (SMEE) \cite{2012PhRvD..86d4023L, 2016PhRvD..94l3012P, 2017PhRvD..96l3013P, 2019PhRvD..99f3018R} uses Principal Component Analysis (PCA) and Bayesian model selection to determine the CCSN explosion mechanism. When PCA is applied to the set of training waveforms, it produces Principal Components (PCs) which represent the most common features of the waveforms in the training set. A linear combination of the first few PCs can then be used as the signal model for each mechanism. By applying PCA to the waveforms, the data can be factored so that 
\begin{equation}
D = U \Sigma V^{T}
\end{equation}
where $D$ is a matrix containing the original waveforms, $U$ and $V$ are matrices whose columns consist of the eigenvectors of $DD^{T}$ and $D^{T}D$, respectively, and $\Sigma$ is a diagonal matrix with elements that correspond to the square root of the eigenvalues of matrix $D$. The $U$ matrix contains the PCs. The main features of the waveforms are contained in just the first few PCs. Each waveform $h_i$ in the data set can be reconstructed using a linear combination of the PCs, multiplied by their corresponding PC coefficients $\beta = \Sigma V^{T}$, 
\begin{equation}
h_{i} = A \sum_{j=1}^{k} U_{j} \beta_{j}
\end{equation}
where $A$ is the amplitude of the signal. The priors on the $\beta$ coefficients are determined by taking the dot product between the PCs and the original waveform matrix. We use a uniform in volume prior on the amplitude, and we assume the sky positions of the CCSN signals are known. We analyse 2\,s of data for each CCSN waveform injection.  

After PCA is used to create the signal models, we then apply the Bayesian model selection. Previous SMEE studies used Bayesian model selection and nested sampling in MATLAB \cite{2012PhRvD..86d4023L} and then with C \cite{2016PhRvD..94l3012P, 2017PhRvD..96l3013P, 2019PhRvD..99f3018R}. In this study, we add the PCA signal model to the python parameter estimation and model selection code Bilby \cite{2019ApJS..241...27A}. We use the dynasty sampler within the Bilby framework. 

We make our PCs in the time domain, using the noise-free training waveforms, and use the standard Bilby gravitational-wave likelihood function described in \cite{2019ApJS..241...27A}. To select the number of PCs we employ the explained variance method. We choose the number of PCs required to represent 80\% of the total variance of the data set. This corresponds to 5\,PCs for the neutrino-driven explosion mechanism, 4\,PCs for the non-exploding models, 3\,PCs for the magneto-rotational explosion mechanism models, and 3\,PCs for the chirplet model. 

The data is whitened by Bilby before we begin the analysis. For LIGO and ET, we carry out our analysis with a minimum frequency of 30\,Hz and a maximum frequency of 1900\,Hz. For the NEMO detector, we use a minimum frequency of 100\,Hz instead, due to the poor low-frequency sensitivity for NEMO.

\subsection{Dictionary Learning}

The Dictionary Learning pipeline is based on the concept of a dictionary, i.e, a matrix of prototype signals stored as columns, also called atoms. By linearly combining their atoms with the adequate conditions, dictionaries can be used for two purposes, denoising and classification. The pipeline used in this work uses both types of dictionaries to obtain optimum classification results. The algorithm 
serves as a continuation of the initial work presented at \cite{llorens-monteagudo_classification_2019, 2022MNRAS.tmp..710S}, along with \cite{2020PhRvD.102b3011T, 2016PhRvD..94l4040T}.

The denoising dictionary $D \in \mathbb{R}^{n\times m}$ is an overcomplete matrix, i.e. the number of atoms $m$ is greater than their length $n$, where the atoms are random fragments of waveforms of the training set. It is applied to the whitened data to recover most of the original waveform from the rest of the data before performing the classification.
Assuming that gravitational waves $u$ are embedded into the detector's noise $n$ following the linear degradation model $f = u + n$\,, the reconstructed waveform can be expressed as a linear combination of atoms of the dictionary, $u \sim D\alpha$\,. Such a combination is found enforcing the coefficients vector $\alpha$ to be sparse while producing a reconstruction as close to the original signal $f$ as possible. This translates into solving the LASSO problem~\cite{lasso},
\begin{equation}
\alpha = \argmin_{\alpha} \left\{ 
  \Vert f - D\alpha \Vert^2_2
  + \lambda \Vert\alpha\Vert_1
\right\}\,,
\end{equation}
where $\lambda$ is the regularization (hyper-)parameter that 
balances the importance between the fidelity term $\Vert
f - D\alpha \Vert^2_2$ and the sparsity of $\alpha$. Its 
value is optimized empirically to produce the 
reconstructions of the injected training set closest to 
their original clean waveforms.
The quality of the reconstructions can be further improved by training the dictionary over the set of training clean data (without background noise), which modifies the dictionary iteratively to become a better representation of the whole training data set. This means solving the double minimization problem
\begin{equation} \label{eq:dictionary-learning}
D = \argmin_{\alpha,D} \frac{1}{n} \sum_{i=1}^p \left\{
  \Vert u_i - D\alpha_i \Vert_2^2 + \lambda\Vert\alpha_i\Vert_1
\right\}\,,
\end{equation}
using the algorithm proposed by Mairal et al.~\cite{Mairal2009}.

While the denoising is a powerful tool to get rid of most of the background noise and improve results, it might limit some methods of data augmentation, like injecting the same waveform into different backgrounds at high SNR.
In this regard, it is worth noting that after applying the denoising the majority of the copies of the test data set were seen by this pipeline as the same signal, thus decreasing the statistical significance of the classification results. As an example, if a waveform was injected 10 times with different (but high enough) SNR, and the original one happened to be misclassified, all 10 of them would potentially be misclassified.

The second and last step is to classify the denoised waveforms using the second kind of dictionary learning technique, named LRSDL by \cite{vu_fast_2017}, specifically developed for classification purposes. It generalises the well-known Fisher discrimination dictionary learning (FDDL)\cite{yang_fisher_2011} by also characterising shared components between different classes of waveforms. This method was shown to perform well in conditions where the available training samples are scarce, and outperform other dictionary-based algorithms in terms of computational resources.
The structure of the LRSDL dictionary is more complex than that of the denoising dictionary. It consists on two matrices corresponding with the class-specific $D$ and shared parts $D_0$, respectively. Given an input waveform $Y$, it tries to replicate it by linearly combining atoms in both parts such that
\begin{equation}
Y \approx DX + D_0 X^0 \,,
\end{equation}
imposing different restrictions to each part of the dictionary. For the class-specific it imposes the FDDL constraints, and for the shared part it enforces $D_0$ to be low-rank. This results in a total of six hyper-parameters; half of them are the physical size of the matrices $D$ and $D_0$, and the other half are regularization parameters which have been optimized for classification after setting the values of the first ones.
Finally, LRSDL by default will always classify an input waveform into one of the known classes regardless of how different it might be. It is not able to tell when a waveform does not belong to any of the known classes. In order to overcome this, as a first approach an empirical threshold has been set at the final value of the loss function of the dictionary. Above this threshold the dictionary is considered to not be confident enough, and the input signal is marked as foreign (unclassified). In this work, this corresponds to the fourth category of waveforms, the chirplets.

We use two different training data sets to train our pipeline. The first one is applied to the denoising dictionary, focused at producing the best reconstruction of any gravitational wave like transient, which means distinguishing the \textit{original} signal from the detector's background noise. The second one is applied to the classification dictionary, hence focused at optimizing the classification of the explosion mechanism by the morphology of the reconstructed waveform.

The first training data set consists of a single collection of all gravitational waveforms without noise (that is, all morphologies except the chirplets). From this, we select 10,000 random windows of 2048 samples (0.5 s) containing at least 512 non-zero samples of strain, which are used to initialize and learn the denoising dictionary as described in eq. \eqref{eq:dictionary-learning}.

The second training data set is divided into our 4 classes (comprising 3 explosion mechanisms and chirplets). For each detector, all waveforms are injected into their respective background noise frame multiple times; at different SNR values (as described in section \ref{sec:noise}), and at two different GPS times. Finally, they are whitened using an averaged estimation of the ASD, but without applying any frequency filter. This approach, diverging from the other pipelines, is aimed to make this pipeline as agnostic as possible to the detector's sensitivity, thus eliminating the need to find specific values of dictionaries' hyper-parameters for each new detector.

\subsection{Convolutional Neural Networks}

The 2D Convolutional Neural Networks (CNN) pipeline, previously implemented in \cite{Iess_2020} and \cite{Iess2023}, classifies signals into a CCSN explosion category based on their characteristics in the time-frequency domain. The CNN model achieves this through a supervised learning procedure, which allows it to catch relevant signal features by applying convolution and pooling operations to the data and produce feature maps. Stacking convolutional layers allows the learning of features on different scales. 

To produce our training data set, we whiten the gravitational-wave strain data in the time domain using the Wavelet Detection Filter \cite{Iess_2020}. A band pass filter is applied with a high frequency cut off of 1900\,Hz, and a low frequency cut off of 30\,Hz for LIGO and ET and 100\,Hz for NEMO. This operation allows the reduction  of the amplitude of the strain data in the frequency regions dominated by the noise background. 
After whitening, we build and store spectrograms of the CCSN signals injected into the noise background choosing a custom resolution in time and frequency. The 2D CNN model is composed of three convolutional layers composed respectively of 32, 64 and 128 filters with kernel sizes $3\times3$, each followed by a max pooling layer. We use the same padding for 
the convolutions. We apply global average pooling after the last convolutional layers, then we flatten the output and pass it through a fully connected layer with 128 nodes. The output of the fully connected layer is passed to a final layer with a softmax activation function to output classification probabilities for the four CCSN explosion mechanism classes. 

The training data is produced by injecting the associated CCSN waveforms into the detectors simulated  background noise,  
with SNRs in the range of 10 to 50. 
We separate $15\%$ of the training data for validation. We train the model by minimizing the cross-categorical loss function. For the process we use the Adam optimizer with an initial learning rate of 0.001. The learning rate is reduced by a factor 0.5 if the validation loss stalls into a plateau for 10 epochs. 

\section{Detector noise and injections}
\label{sec:noise}

\begin{figure}
\includegraphics[width=\columnwidth]{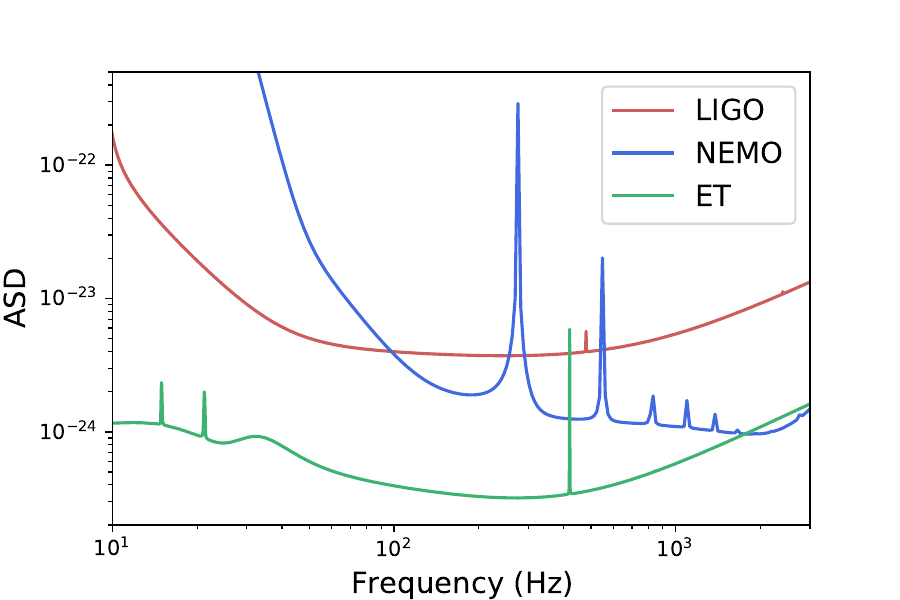}
\caption{The amplitude spectral density (ASD) curves for the Advanced LIGO design sensitivity \cite{2015CQGra..32g4001L},  Einstein Telescope \cite{2011CQGra..28i4013H} and NEMO \cite{2020PASA...37...47A} gravitational-wave detectors.
Einstein Telescope has a similar frequency band to LIGO but it is more sensitive. NEMO has good high frequency sensitivity, but it has poor sensitivity at low frequencies. }
\label{fig:asd}
\end{figure}

We use 4096\,s of data from the LIGO Livingston detector during the third observing run, starting at GPS time 1238179840. The data was downloaded from the Gravitational Wave Open Science Center (GWOSC) \cite{2021SoftX..1300658A}. The sample rate is 4096\,Hz. For the LIGO and ET detectors, we recolour the real LIGO detector noise to the detectors expected design sensitivities. We do this so that the impact of real noise transients that occur in the data will be included in our study. We use simulated Gaussian noise for the NEMO detector. This is because of the significant differences in the frequency sensitivity between the real LIGO observing run noise and the NEMO frequency band. The noise curves used to produce the detector noise are shown in Figure \ref{fig:asd}. For ET, we use the ET-D configuration \cite{2011CQGra..28i4013H}.

After creating the noise for each detector, we add the signals in 10 second intervals into the data. In total there are 368 signals in each detector. There are a total of 92 non-exploding signals, 138 neutrino-driven explosions, 92 magneto-rotational explosions and 46 chirplets. Before adding the signals into the noise, we resample them to 4096\,Hz, we filter out the frequencies below 30\,Hz for LIGO and ET, and 100\,Hz for NEMO, and scale the amplitude for the signal to noise ratio (SNR) that we require. We use SNR values of 25, 30, 35, 40, and 45, as a reasonably high SNR is required for CCSN signals to be detected by gravitational-wave burst searches \cite{2021PhRvD.104j2002S}.

\section{Results}
\label{sec:results}

\subsection{Advanced LIGO}

\begin{figure*}
\includegraphics[width=5.8cm]{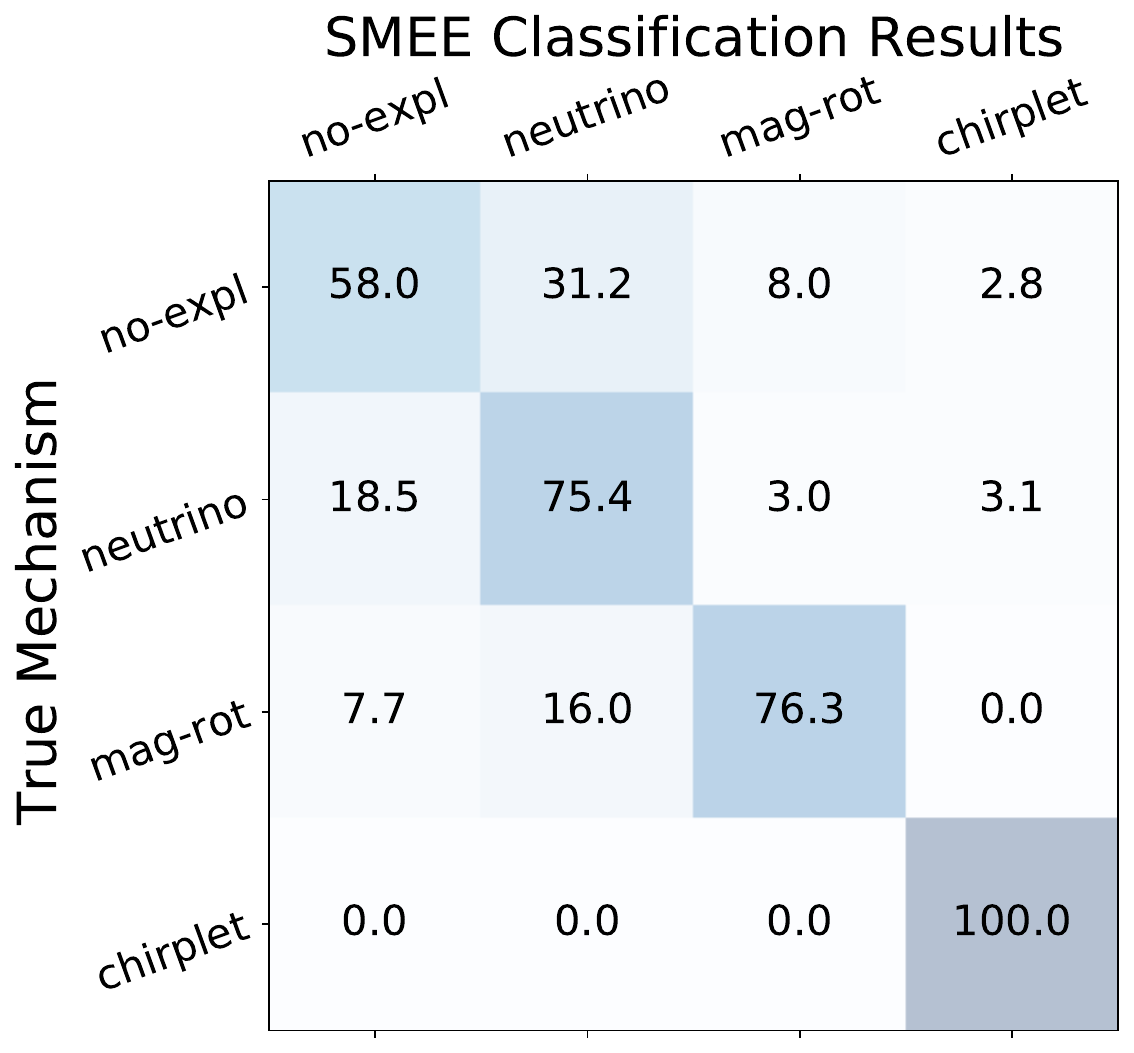}
\includegraphics[width=5.8cm]{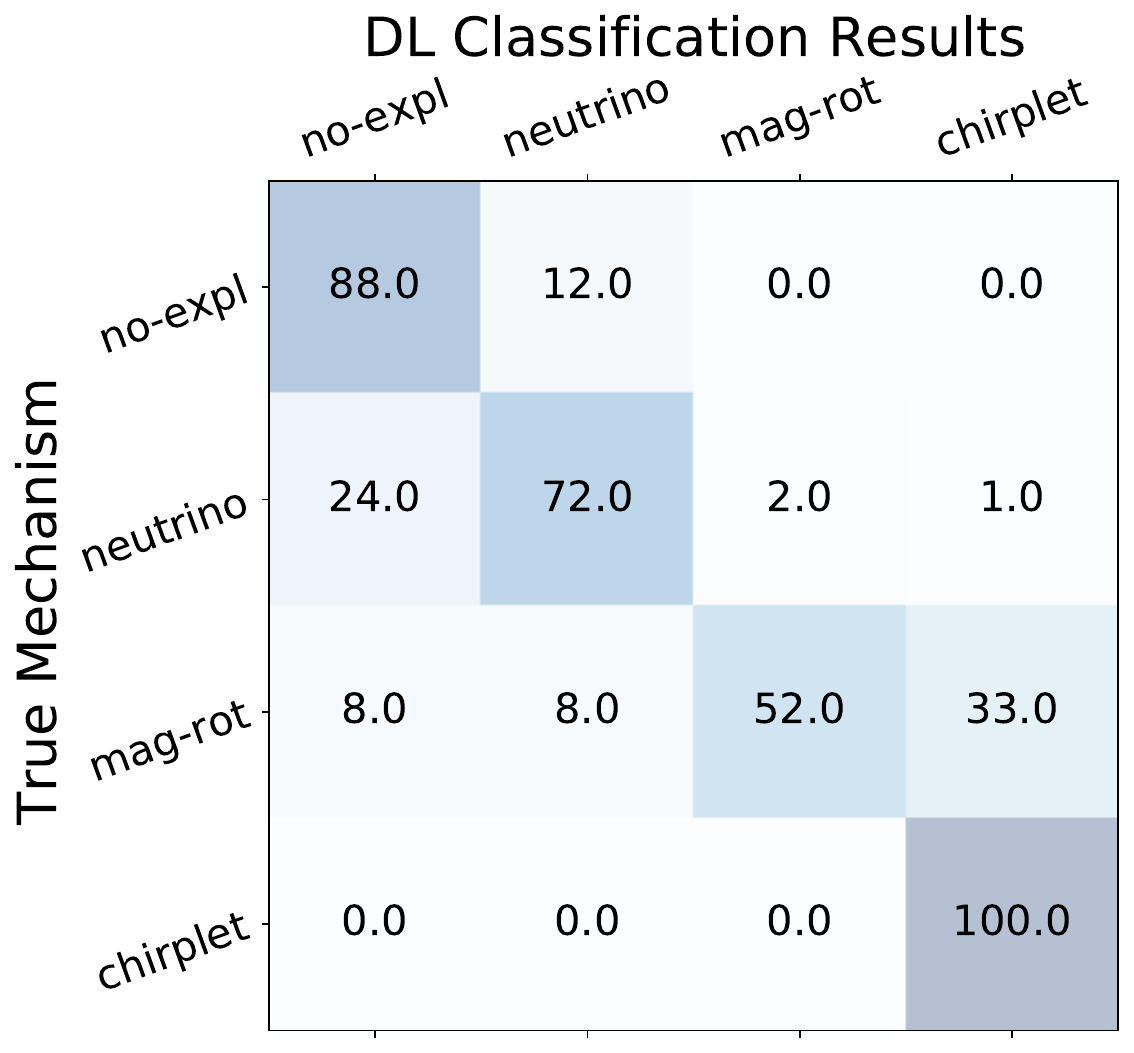}
\includegraphics[width=5.8cm]{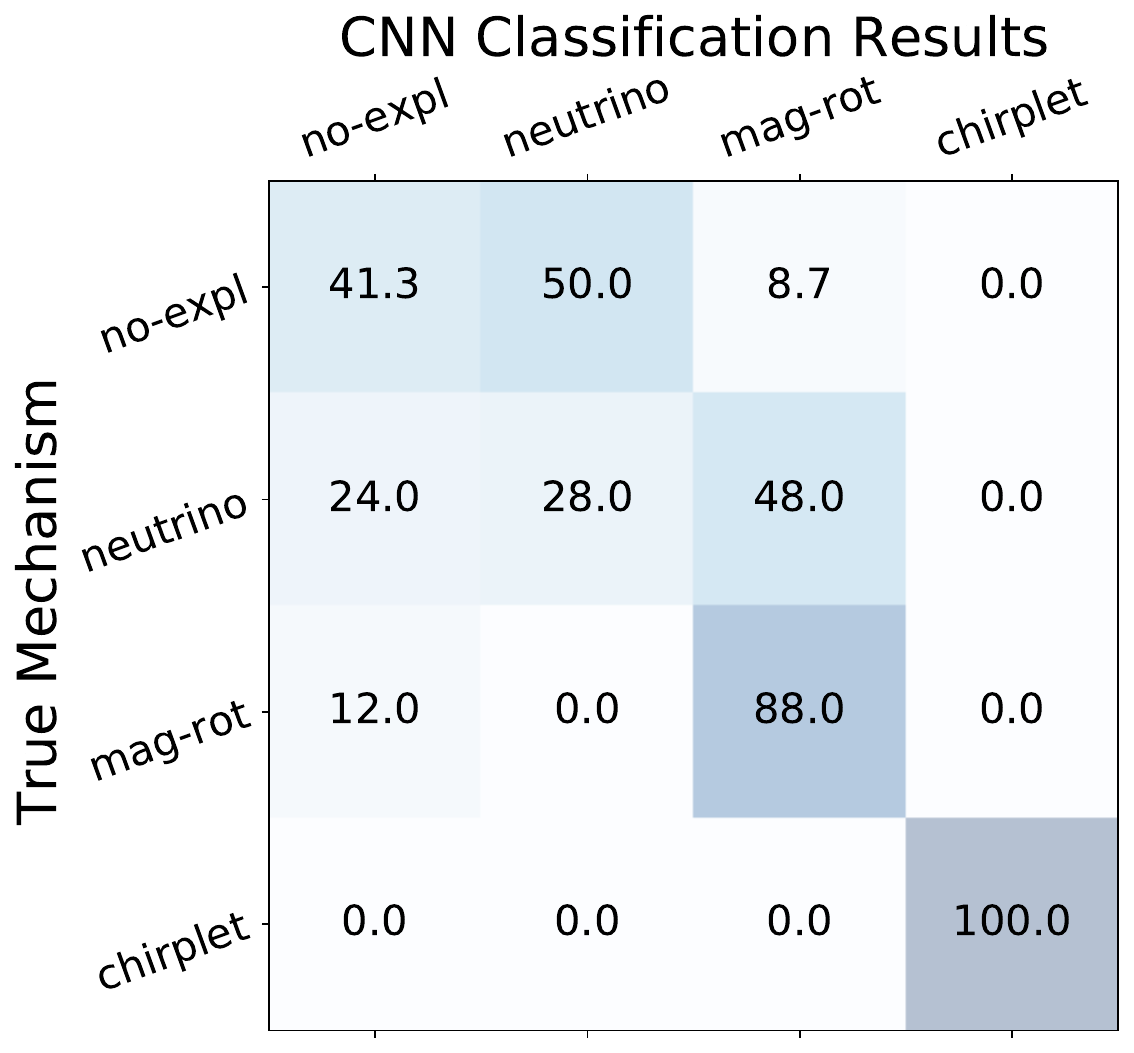}
\caption{ The classification results for the LIGO detector. From left to right are the results for Bayesian model selection, Dictionary Learning, and 2D CNN. The Bayesian model selection and 2D CNN methods were able to correctly classify every chirplet signal. The Dictionary Learning method produces the best results for the non-exploding models.  
}
\label{fig:ligo_matrix}
\end{figure*}

The results for all the waveforms injected into the LIGO detector noise are shown in Figure \ref{fig:ligo_matrix}. All methods were able to classify 100\% of the chirplets correctly. This shows that our techniques are able to tell when a similar gravitational-wave signal does not fit into one of our CCSNe explosion mechanism categories. 

The magneto-rotational explosion mechanism signals were correctly classified with a high accuracy by the Bayesian model selection and the 2D CNN method, over 75\%. The dictionary learning method only classified 52\% of the magneto-rotational explosion mechanism signals correctly, and the majority of the rest of the signals were mixed into the chirplet class. 

The neutrino-driven explosion models were classified with over a 70\% accuracy by the Bayesian model selection and dictionary learning methods. This shows that the two methods are able to correctly determine that the m39 rapidly-rotating model was a neutrino-driven explosion even though the rapid rotation results in very similar features to the magneto-rotational explosion models. This was not the case for the 2D CNN method, as 48\% of the neutrino-driven explosion waveforms were classified into the magneto-rotational waveform class. 

For the non-exploding waveform class, the dictionary learning method has the best results with 88\% of the waveforms correctly classified. For the Bayesian model selection method, the non-exploding waveforms that were incorrectly classified were mainly mixed in with the neutrino-driven explosions. Those signals were the ones with the lowest SNR, where it is more difficult to detect the lower amplitude SASI components of the gravitational-wave signal. The 2D CNN classifies half of the non-exploding waveforms into the neutrino-driven explosion class. 

\subsection{Einstein Telescope}

\begin{figure*}
\includegraphics[width=5.8cm]{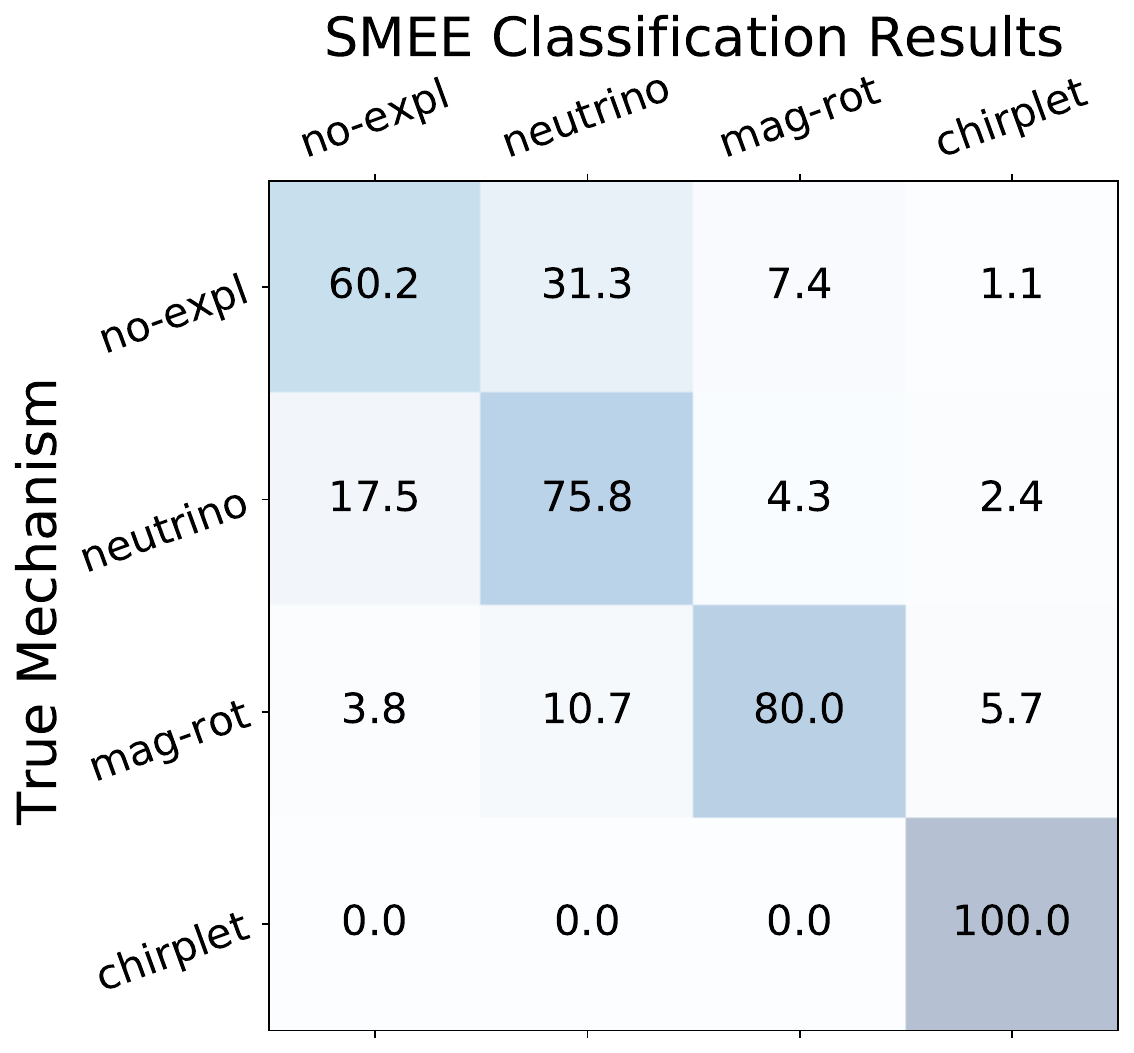}
\includegraphics[width=5.8cm]{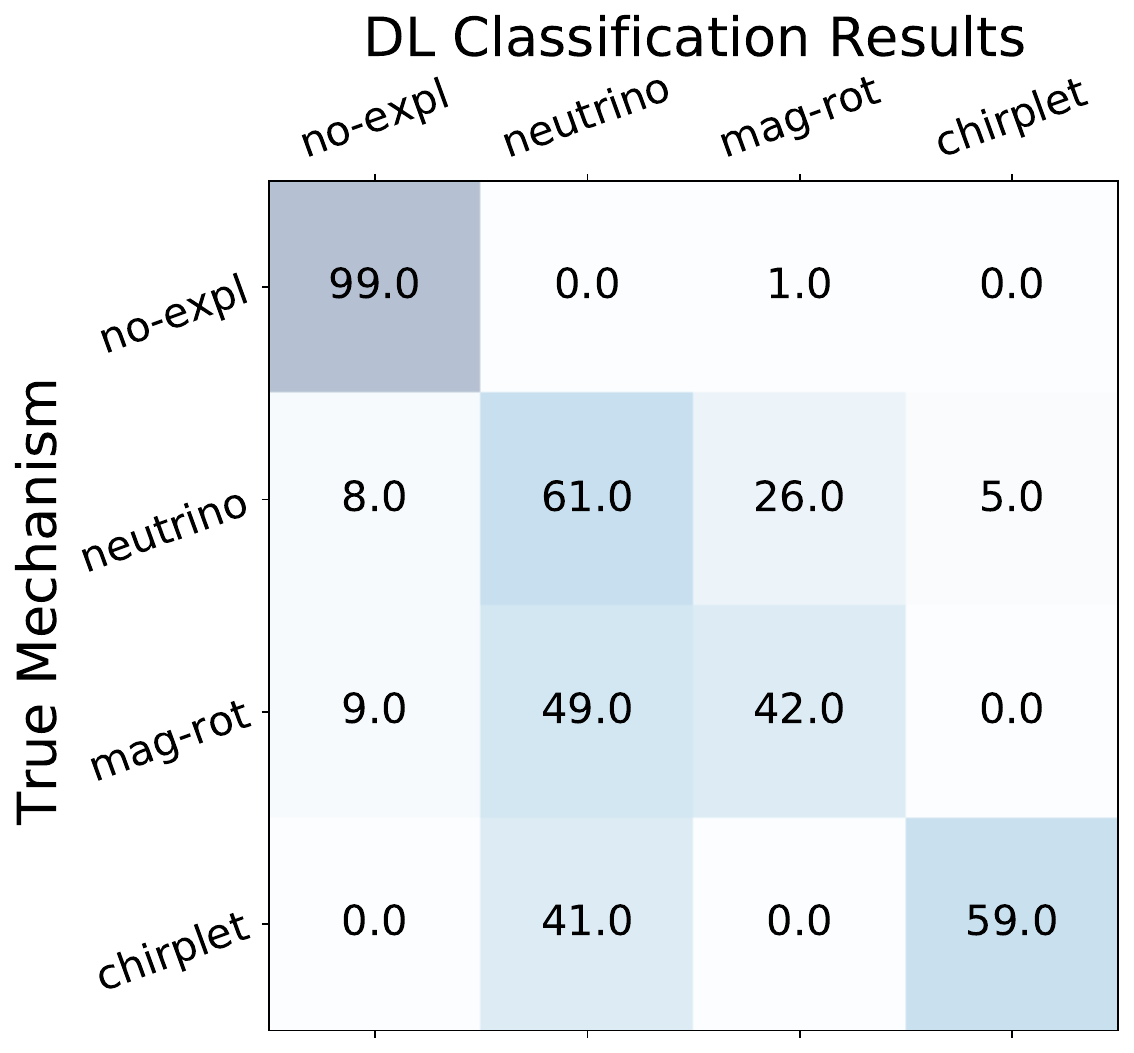}
\includegraphics[width=5.8cm]{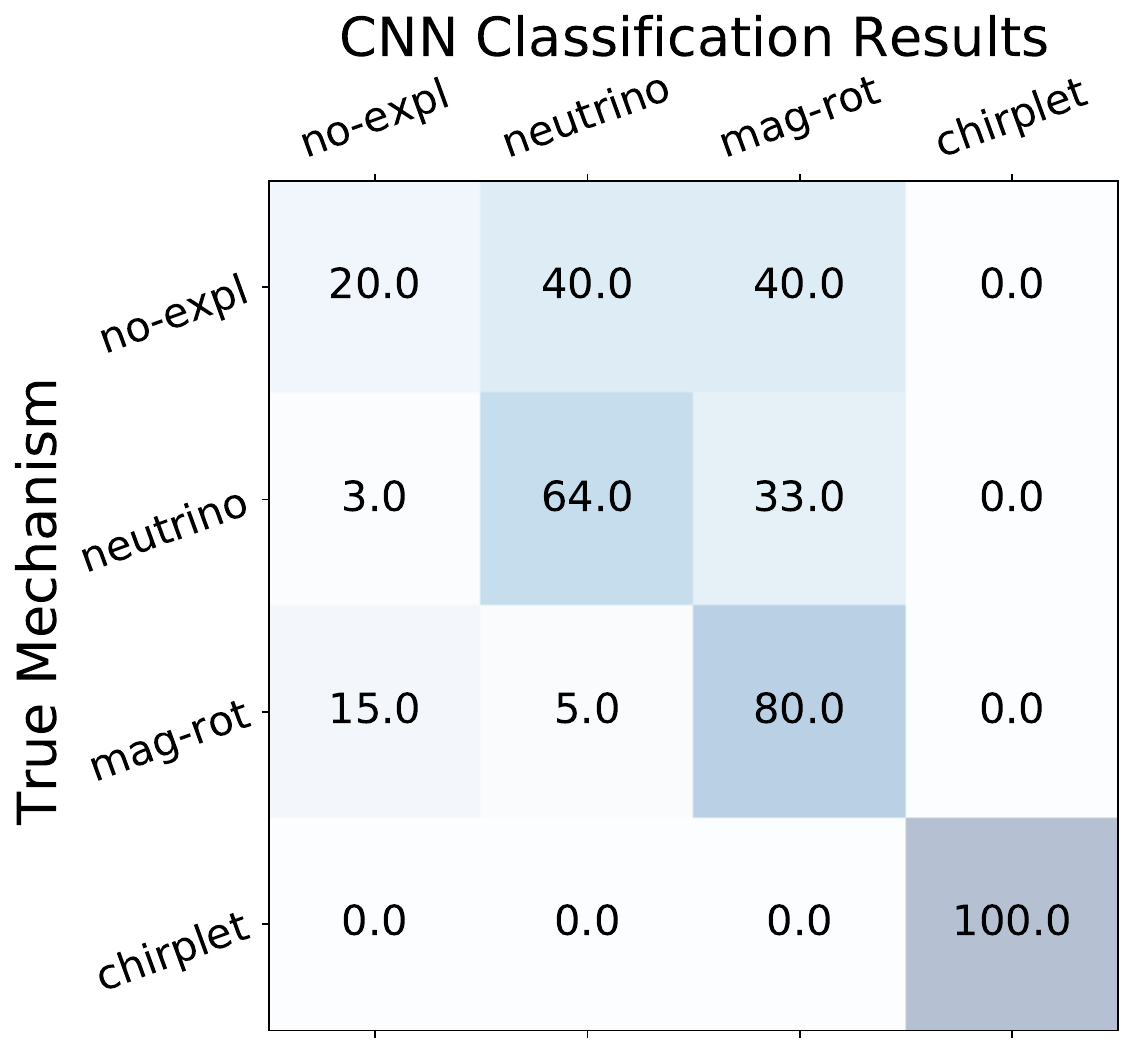}
\caption{The classification results for the ET detector. From left to right are the results for Bayesian model selection, Dictionary Learning, and 2D CNN. The results for all three methods are similar to those obtained for Advanced LIGO. }
\label{fig:ET_matrix}
\end{figure*}

The results for the ET detector, shown in Figure \ref{fig:ET_matrix}, are similar to those obtained for Advanced LIGO. For the Bayesian model selection method, the accuracy is slightly better for ET. However, there is still some mixing between the non-exploding and neutrino-driven explosion classes.

For the Dictionary Learning method, almost every non-exploding waveform is correctly classified. This increase in accuracy may be due to the increased low frequency sensitivity of the ET detector. However, the accuracy is reduced for the neutrino, magneto-rotational and chirplet classes. This may be due to the pipeline's hyper-parameters being optimised using the LIGO detector noise. There is a large computational effort required to redo the optimisation of the hyper-parameters for different detectors noise, but our results show that retraining on the latest noise data may be necessary to obtain the best results for a real gravitational-wave detector. 

The 2D CNN results for ET are also similar to those obtained for Advanced LIGO. The 2D CNN classifies the chirplet and magneto-rotational models with a high accuracy, however there is a lot of mixing between classes for the neutrino-driven explosions and non-exploding models. 

\subsection{NEMO}

\begin{figure*}
\includegraphics[width=5.8cm]{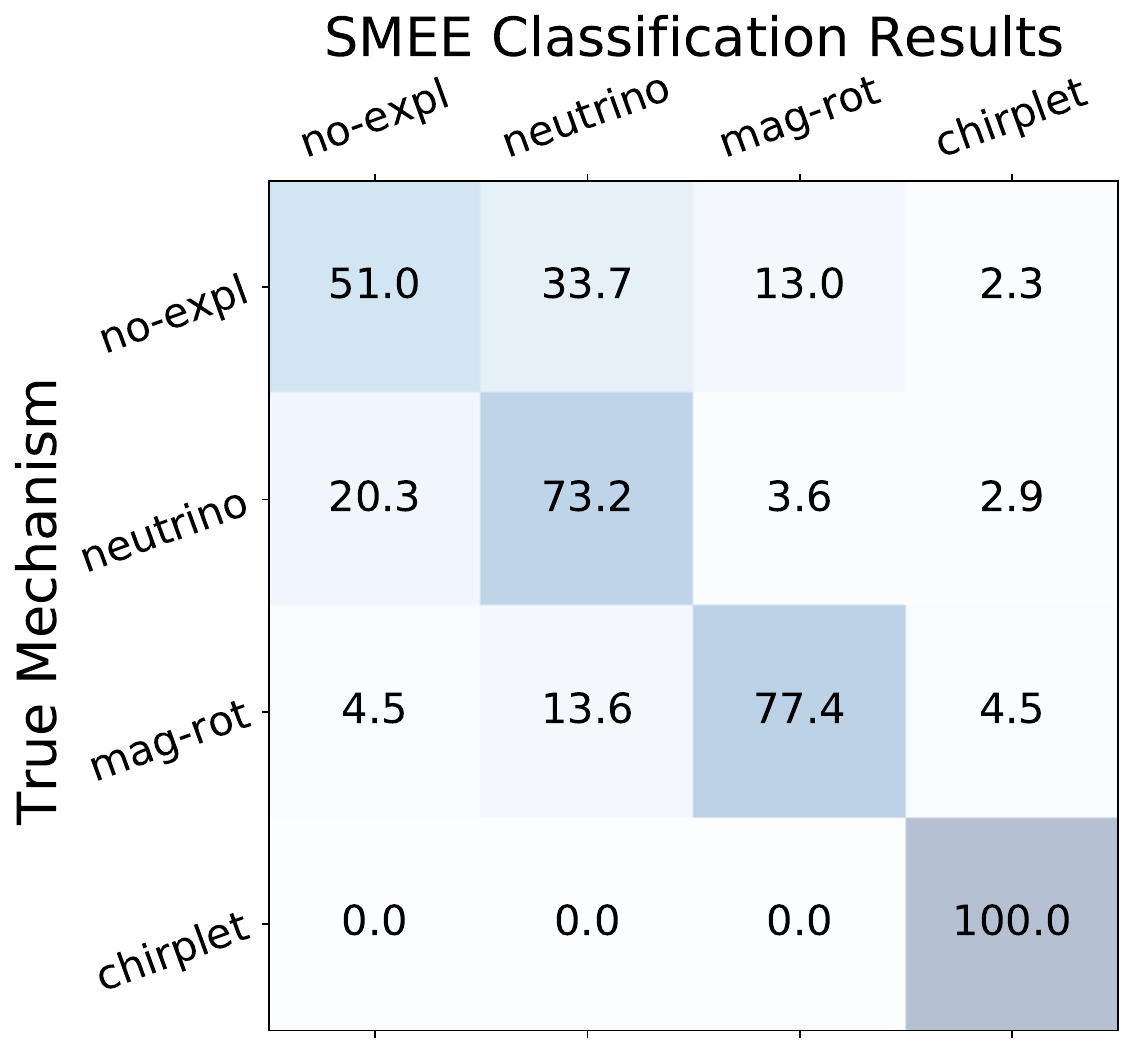}
\includegraphics[width=5.8cm]{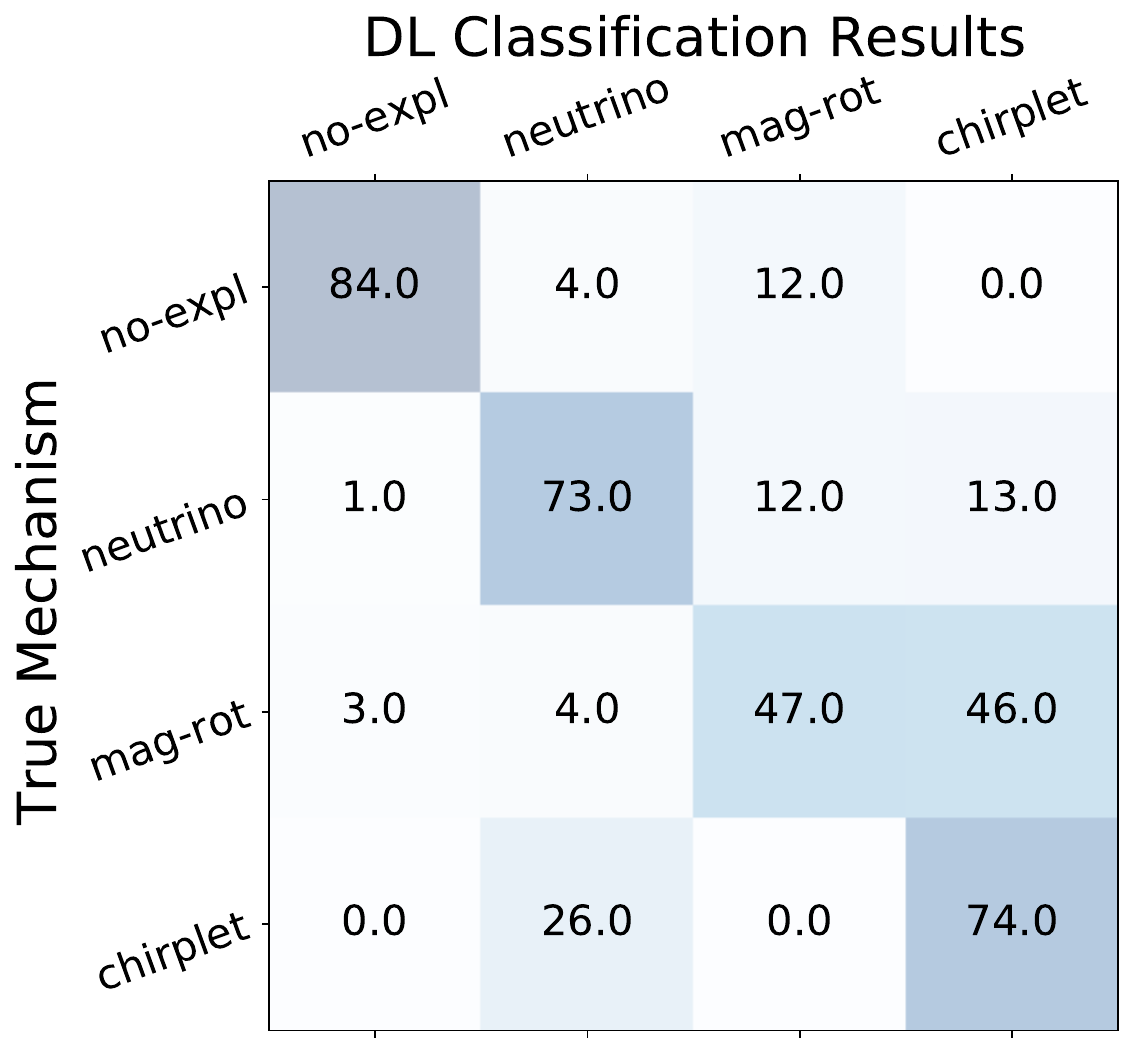}
\includegraphics[width=5.8cm]{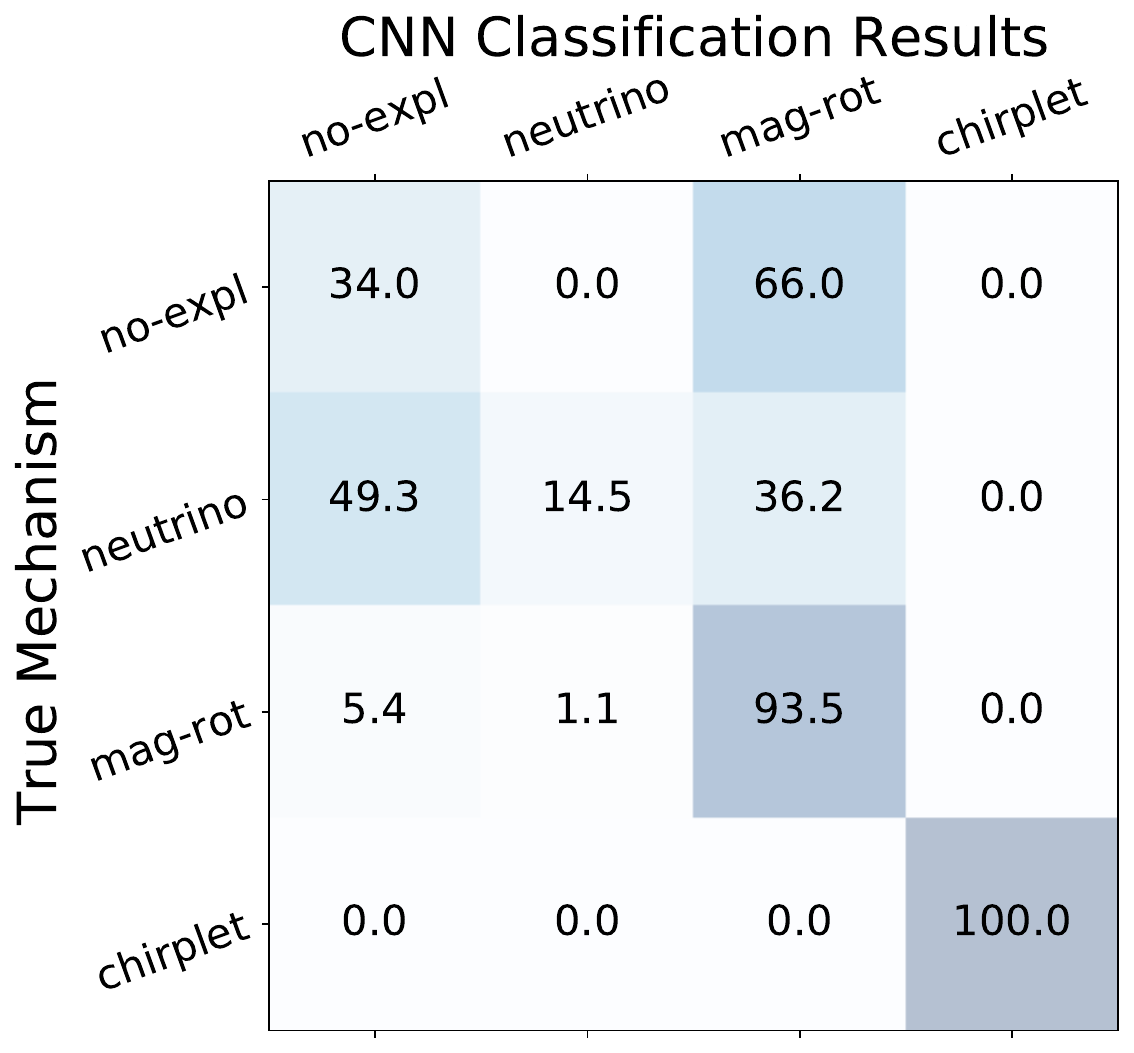}
\caption{The classification results for the NEMO detector. From left to right are the results for Bayesian model selection, Dictionary Learning, and 2D CNN. In the NEMO detector, the magneto-rotational waveforms and the chirplets are classified with a high accuracy. It is more difficult to distinguish between neutrino and non-explosions as the SASI modes are not visible due to the detectors poor low frequency sensitivity. }
\label{fig:NEMO_matrix}
\end{figure*}

The results for the NEMO detector are shown in Figure \ref{fig:NEMO_matrix}. For the Bayesian model selection method, the chirplet signals and the magneto-rotational signals are classified with a high accuracy. The accuracy for the non-exploding class is reduced for NEMO. This is likely due to the poor low frequency sensitivity of the NEMO detector. The main difference between the neutrino-driven and non-exploding models is the longer and stronger low frequency SASI component in the non-exploding models, but this part of the signal is not as visible in the NEMO frequency band. 

The NEMO results for the Dictionary Learning method are similar to those for ET and LIGO. A large number of the magneto-rotational waveforms are incorrectly classified (about 53\%). However the chirplet, neutrino-driven and non-exploding waveforms are all classified with a high accuracy. 

The 2D CNN method is able to classify the chirplet and magneto-rotational waveforms with a high accuracy (95.3\% and 100\%, respectively). However, as for the other two detectors, most of the neutrino-driven and non-exploding waveforms are mixed between the different classes.

\section{Conclusions}
\label{sec:conclusion}

A gravitational-wave detection from a CCSN may enable us to determine the mechanism that powered the explosion. Currently, most typical CCSN explosions are thought to be powered by absorption of some of the energy from neutrinos. The explosion of more energetic CCSNe are expected to be powered by the combined effects of rotation and magnetic fields. 

Previous studies have tried to determine if we can use a gravitational-wave detection to distinguish between these two explosion mechanisms \cite{2012PhRvD..86d4023L, 2016PhRvD..94l3012P, 2017PhRvD..96l3013P, 2022MNRAS.tmp..710S}. 
However, at the time of these previous studies, there were no long duration waveforms available for the magneto-rotational explosion mechanism. Waveforms were assumed to be magneto-rotational explosions if they were rapidly-rotating. However, it is possible for a rapidly-rotating star to still undergo a neutrino-driven explosion.  
Therefore, in this paper we have carried out an updated study that uses long duration gravitational waveforms in which rotation and magnetic fields play either a supporting role or a leading role in driving the explosion. 
We have also added a non-exploding category to our mechanism classifications. Non-exploding models are very similar to neutrino-driven explosions, but they usually display some extra features in spectrograms, such as longer duration low-frequency mode from the SASI. It is possible that some models that were only simulated for a short duration may have undergone a neutrino-driven explosion eventually if they were continued for a longer time. 
The boundary between non-exploding and exploding stars is further blurred by the fact that black hole formation can occur not only in failed CCSNe, but also may be preceded by a neutrino-driven explosion \cite{2021MNRAS.503.2108P}.

We have used noise for three different gravitational-wave detectors: Advanced LIGO, Einstein Telescope and NEMO with noise levels given by the expected design sensitivities. The Advanced LIGO and ET detectors have a similar frequency band, but the ET detector is significantly more sensitive. The NEMO detector has a better high frequency sensitivity than the other gravitational-wave detectors. We have injected a total of 368 CCSN gravitational-wave signals into the noise of each detector with SNR values between 25 and 45. 
We have then classified all of the injected signals with three different methods that have been used previously for the classification of CCSN signals. They are Bayesian model selection, Dictionary Learning and a 2D CNN. 

The results have shown that, regardless of the method, we are able to distinguish between rapidly-rotating neutrino-driven explosions and magneto-rotational explosions, and in general we can distinguish our new long-duration magneto-rotational explosion waveforms from the other types of CCSN explosion models. We have also found that we can accurately classify chirplet signals as not being from any one of our explosion mechanisms.
Distinguishing between neutrino-driven explosions and non-exploding models is more difficult, as expected. The non-exploding models are expected to have a more prolonged SASI signal component than most neutrino-driven explosion models, however this is only visible at high SNR. For the LIGO and ET detectors, correctly classifying a waveform as being non-exploding is only possible when the SNR is high and the SASI component in the gravitational-wave signal is clearly visible. For the NEMO detector it is even more difficult, due to the detector's poor sensitivity in the low-frequency band where the SASI produces the most gravitational-wave energy. 
In the future, the results could be improved for the ET detector by including the memory component of the gravitational-wave signal, which is only present in exploding models. The memory component of the gravitational-wave signal occurs at frequecies too low for current ground based gravitational-wave detectors and NEMO.

In the event of a CCSNe occurring in the near future, we would be able to use every waveform available to us for the training, as we wouldn't need to keep some spare for testing. Having the extra waveforms available would increase our accuracy, and we will also include any new waveforms that are continuously being produced in numerical simulations. The magneto-rotational explosion mechanism still only has a small number of full 3D waveforms available. As more sensitive gravitational-wave detectors become available, our sensitivity to CCSNe signals will increase. However the increased sensitivity will also result in extra data analysis challenges that will need to be addressed in future work, for example the CCSN overlapping with a binary black hole signal. 

\begin{acknowledgments}
JP is supported by the Australian Research Council's (ARC) Discovery Early Career Researcher Award (DECRA) project number DE210101050, and the ARC Centre of Excellence for Gravitational Wave Discovery (OzGrav) project number CE170100004. BM acknowledges support by ARC Future Fellowship FT160100035. MLlM, ATF, MO and JAF acknowledge support from the Spanish Agencia Estatal de  Investigaci\'on  (Grants PID2021-125485NB-C21 and PID2021-127495NB-I00, funded by MCIN/AEI/10.13039/501100011033 and ERDF A way of making Europe), by the Generalitat Valenciana (Prometeo grants CIPROM/2022/13 and CIPROM/2022/49), and by the Astrophysics and High Energy Physics programme of the Generalitat Valenciana ASFAE/2022/03 and ASFAE/2022/026 funded by MCIN and the European Union NextGenerationEU (PRTR-C17.I1). Further support is provided by the EU's Horizon 2020 research and innovation (RISE) programme H2020-MSCA-RISE-2017 (FunFiCO-777740), and  by  the  European Horizon  Europe  staff  exchange  (SE)  programme HORIZON-MSCA-2021-SE-01 (NewFunFiCO-101086251). MO acknowledges support from the Spanish Agencia Estatal de  Investigaci\'on via the Ram\'on y Cajal programme (RYC2018-024938-I). AI is supported by the
by the European Union Horizon Programme call INFRAEOSC-03-2020 - Grant Agreement Number 101017536. This material is based upon work supported by NSF's LIGO Laboratory which is a major facility fully funded by the National Science Foundation. 
This work has used the following open-source packages: \textsc{NumPy} \citep{harris:2020}, \textsc{SciPy} \citep{scipy:2020} and \textsc{Matplotlib} \citep{Hunter:2007}.

\end{acknowledgments}


\bibliographystyle{apsrev}

\bibliography{bibfile}

\end{document}